# Spin diffusion length and polarization of ferromagnetic metals measured by spin-absorption technique in lateral spin valves

G. Zahnd[1], L. Vila[1,*], V. T. Pham[1], M. Cosset-Cheneau[1], W. Lim[1], A. Brenac[1], P. Laczkowski[1], A. Marty[1], and J. P. Attané[1,†]

[1] Univ. Grenoble Alpes, Institute for Nanoscience and Cryogenics, CEA, CNRS, SPINTEC, F-38000 Grenoble, France

**ABSTRACT**

We present measurements of pure spin current absorption on lateral spin valves. By varying the width of the absorber we demonstrate that spin current absorption measurements enable to characterize efficiently the spin transport properties of ferromagnetic elements. The analytical model used to describe the measurement takes into account the polarization of the absorber. The analysis of the measurements allows thus determining the polarization and the spin diffusion length of a studied material independently, contrarily to most experiments based on lateral spin valves where those values are entangled. We report the spin transport parameters of some of the most important materials used in spinorbitronics ($Co_{60}Fe_{40}$, $Ni_{81}Fe_{19}$, Co, Pt, and Ta), at room and low (10 K) temperatures.

**TEXT**

Recent advances in spinorbitronics have evidenced the need for new characterization techniques, measuring precisely the spin-dependent material properties. Indeed, parameters such as the damping, the polarization, the spin diffusion length, the Dzyaloshinskii-Moriya interaction (DMI) or the spin Hall angle are key factors to understand and control various mechanisms: spin transfer torque[1,2,3], spin texture dues to DMI[4,5,6], or charge-to-spin conversion based on spin orbit coupling effects at Rashba interfaces[7,8], in the bulk of SHE materials[9,10], or in topological insulators[11].

Among those basic material parameters, the spin diffusion length occupies a primordial role in all spin transport mechanisms. Pioneer spin transport measurements of the spin diffusion length[12,13,14] have been performed on vertical structures, using thickness dependences, but the data available on such material parameters determination are often restricted to low temperatures. Almost two decades later, the determination of short spin diffusion lengths remains difficult, for instance in ferromagnetic alloys and heavy metals. The need for precise measurement techniques has however become crucial, in particular in spinorbitronics: the spin diffusion length is central in the controversies concerning the determination of the charge-to-spin conversion rates[15].

* Laurent.vila@cea.fr
† Jean-philippe.attane@cea.fr

Gap dependence measurements in lateral spin-valves[16,17] (LSV) are well adapted for materials with long spin diffusion lengths. For materials with short spin diffusion length, spin-pumping measurements with thickness dependences have been used to study heavy metals with strong spin-orbit coupling[18,19], but even for Pt, which is the most studied SHE material, the extracted values displayed in the literature are widely scattered[15,20,21,22,23].

Spin diffusion lengths of heavy metals have also been extracted using non-local measurements in LSVs[24,25,26]. The method consists in adding a wire with a small spin diffusion length in-between the ferromagnetic electrodes of a LSV, connected to the conducting channel. The detected spin signal is then found to decrease, because of the absorption of the pure spin-current in the added wire. The absorption process being linked to the spin resistance of the material, it is possible to determine its spin diffusion length.

Importantly, using this technique there is no need for varying the thickness of the studied material. As the spin diffusion length of a material varies with its resistivity[22,27,28], and thus often varies with the thickness, it is difficult to use a thickness dependence to measure the spin diffusion length.

Up to now, spin absorption measurements have been performed to extract the spin diffusion length of non-magnetic heavy materials. As it has been done one NiFe/Cu nanostructures[29], we show that this spin absorption technique is well adapted to study ferromagnetic materials with short spin diffusion lengths[30] and extract their polarization. Indeed, it allows to measure separately the polarization and the spin diffusion length, which cannot be extracted independently in most LSVs-based experiments. We then measure the spin diffusion length of some of the most important materials used in spinorbitronics ($Co_{60}Fe_{40}$, $Ni_{81}Fe_{19}$, Co, Pt, and Ta), at room and low (10 K) temperatures as well as the polarization of the ferromagnetic materials.

The devices nanofabricated and measured in this paper are lateral spin-valves[31], consisting in two ferromagnetic nanowires connected by a perpendicular non-magnetic channel. In all devices but the reference ones, a wire acting as a spin-absorbing element has been inserted between the ferromagnetic electrodes (cf. figures 1a and 1b). The devices have been patterned by e-beam lithography on a $SiO_2$ substrate. The nanowires of spin-absorbing materials have been fabricated by evaporation of pellets through a patterned PMMA resist mask and subsequent lift-off. In a second lithography step, both the ferromagnetic electrodes and the non-magnetic channel have been realized by multiple angle deposition[32]. An argon ion beam milling has been used in order to obtain clean transparent interfaces between the non-magnetic channel and the spin-sink materials.

In the case of ferromagnetic spin-absorbing materials, the electrodes and the absorbing wire are constituted of the same element. Hence, only one step of lithography and multi-angle deposition have been used. All the nanodevices are geometrically identical, except for the widths of the spin-absorbing wires (cf. figure 1c). In the case of non-magnetic absorbing elements, all the ferromagnetic electrodes are made of CoFe.

Cu has been used as nonmagnetic material for the conducting channel, in order to optimize the interface quality and to limit its influence on the spin accumulation relaxation: indeed, Al/CoFe interfaces has been shown to induce resistive interfaces (more details are presented in the supplementary materials). The ferromagnetic and absorbing wires are all 20 nm thick, the non-magnetic channels are 80 nm thick, while ferromagnetic and non-magnetic nanowires are 50 nm wide. The distance center to center between the two ferromagnetic electrodes is 300 nm.

The resistivities were measured by using a Van der Pauw[33] method, and the parameters of Cu are taken from previous measurement and control samples[17]. The resistivity of the Cu channel is found to be of 3.5±0.4 µΩ.cm at 300K and of 2.5±0.3 µΩ.cm at 10K. Its spin diffusion length had been previously determined by a study based on a gap dependence[17,34], and has been determined to be of 350 nm at 300 K and 700 nm at 10 K. The geometric parameters of the devices have been characterized by SEM, and resistivity measurements have been performed for all the studied elements.

Figure 1(a) shows the probing configuration corresponding to classical non-local measurements[35] on lateral spin valves. In this configuration, a current flowing through the ferromagnetic/non-magnetic interface induces a spin accumulation near the interface. The diffusion of majority electrons away from this region, and of minority electrons towards this same region, leads to the generation of a pure spin current flowing in the non-magnetic channel. This spin current relaxes over the spin diffusion length of the conducting material. The voltage measurement at the second ferromagnetic/non-magnetic interface probes the electrochemical splitting of the two spin populations, which corresponds to the local remaining spin accumulation. The reversal of the two magnetizations can be obtained by applying an external field along the easy axis of the ferromagnetic electrodes. The parallel and anti-parallel states correspond to high and low spin signals state[16,28] (cf. fig. 2), the difference between the spin signals, *i.e.* the spin signal amplitude, being proportional to the spin accumulation at the detecting electrode. All the measurements have been performed at 300 K and 10 K, using a lock-in technique with an applied current of 100 µA.

The non-local measurements shown on figure 2 were obtained in devices possessing inserted spin-absorbing Platinum wires of different widths. The upper curve (in grey) corresponds to a reference sample (*i.e.* a classical lateral spin valve, without Pt absorber). When spin-sink wires are inserted, the spin signal amplitude decreases, since a part of the diffusing spin current is absorbed and relaxes in Pt. As the signal decrease is directly linked to the amount of spin current absorbed by the Pt wire[24,25,26], it allows thus to determine the spin diffusion length of Pt.

The spin signal decrease for a given spin diffusion length can be calculated using 1D analytical expressions derived from the Valet Fert model. A description recently proposed[26] assimilate the channel in parallel with the absorbing element as a global material over the contact length

(cf. figure 2b). In this study, we adapt the analytical model to extend the spin absorption study to the case of ferromagnetic elements. Taking into account the polarization of the absorbing material, the described bilayer can be assimilated to an effective material possessing the following spin diffusion length:

$$\lambda_N^* = \frac{\lambda_N}{\sqrt{1+\frac{\rho_N\,(1-P_A^2)}{\rho_A}\frac{\lambda_N^2}{\lambda_A t_N}\tanh(\frac{t_A}{\lambda_A})}} \qquad (1)$$

where $\lambda_i$, $\rho_i$, $P_i$ and $t_i$ are respectively the spin diffusion length, the resistivity, the polarization and the thickness of the $i^{th}$ material. The material $N$ is the non-magnetic channel, while the material $A$ is the absorber. The effective spin diffusion length $\lambda_N^*$ reflects the two-parallel relaxation path, *i.e.* the channel and the absorber, for the out-of-equilibrium spin accumulation. The analytical model used to extract the polarization and the spin diffusion length from the spin signal amplitude is detailed in the supplementary materials.

The spin transport parameters of platinum and tantalum have been extracted (cf. table 1). The spin diffusion length of Ta is found to be of around 2 nm at both 300 K and 10 K. The measured resistivity of Ta being also invariant from temperature (200 µΩ.cm), these results are in good agreement to the assumption that the product $\rho\lambda$ is independent of temperature for most materials[22,27,36].

For Pt, we find a very good agreement with recent spin pumping experiments[15,22], with a spin diffusion length of 3.8 nm for a 20 nm thick Pt wire of 18 µΩ.cm at room temperature, increasing to 4.6 nm at 10 K. These results are also in qualitatively good agreement to the theoretic expectations that $\rho\lambda$ is constant[22,27,28] and to the spin diffusion lengths value reported in [27] for similar Pt resistivities.

Let us now apply this model to the case of ferromagnetic materials. In most of the studies concerning ferromagnetic elements, the estimation of the spin diffusion length $\lambda_A$ and of the polarization $P_A$ is difficult, since they are usually entangled together in a single effective parameter, as in the case of spin resistances.

If the ferromagnetic electrodes are made with a different material than the absorbing wire, then the dependence of the spin signal on the absorber does not allow to determine $\lambda_A$ and $P_A$ independently. Here, we propose a simple solution, consisting in the use of a nanodevice made with the same ferromagnetic material $A$ for the electrodes and the absorbing wire. In that situation, the material parameters take place both in the spin injection and detection efficiency of the electrodes (*i.e.*, the amplitude of the spin signal) and in the absorption efficiency of the spin signal (*i.e.*, the decreasing profile of the spin signal with the width $w_A$). The injection efficiency primarily depends on the polarization $P_F = P_A$, while the absorption efficiency depends on another parameter (more details in annexes and on figure 3e and 3f). Hence, this measurement provides a way to efficiently disentangle the values of $P_F$ and $\lambda_F$.

Devices with CoFe, NiFe and Co electrodes and spin sinks have been studied in order to measure their spin diffusion lengths and polarizations at both 300 K and 10 K. The evolutions of the spin accumulation are presented in figure 3. The dependence of the spin signal as a function of $w_A$ are fitted using the expression presented in the supplementary materials. The obtained parameters for each material are displayed in table 1. The errors bars for the polarization and the spin diffusion length (presented in figure 3 e and f) have been determined by considering the acceptable set of parameters leading to spin signal profiles fitting the experimental point with a good correlation.

*Table 1: Spin transport parameters extracted from the spin absorption experiment for each material, at both room and low (10 K) temperature. The resistivities are measured using a Van der Pauw method.*

|      | ρ (300K) μΩ.cm | P (300K) ø | λ (300K) nm | ρ (10K) μΩ.cm | P (10K) ø | λ (10K) nm |
|------|---|---|---|---|---|---|
| CoFe | $20 \pm 1.3$ | $0.48^{+0.0}_{-0.02}$ | $6.2^{+0.3}_{-0.7}$ | $15 \pm 0.9$ | $0.48^{+0.03}_{-0.01}$ | $8.3^{+0.7}_{-1.8}$ |
| NiFe | $30 \pm 3$ | $0.22^{+0.05}_{-0.06}$ | $5.2^{+1.8}_{-0.9}$ | $22 \pm 1.2$ | $0.40^{+0.1}_{-0.03}$ | $5.8^{+0.2}_{-1.8}$ |
| Co   | $25 \pm 2.4$ | $0.17^{+0.08}_{-0.02}$ | $7.7^{+1.8}_{-2.2}$ | $15 \pm 1.6$ | $0.18^{+0.09}_{-0.03}$ | $12.5^{+3.5}_{-3.7}$ |
| Pt   | $18 \pm 0.7$ | ø | $3.8^{+0.7}_{-0.3}$ | $13 \pm 0.4$ | ø | $4.8^{+0.6}_{-0.5}$ |
| Ta   | $200 \pm 15$ | ø | $1.9^{+0.3}_{-0.5}$ | $200 \pm 15$ | ø | $2.0^{+0.4}_{-0.6}$ |

The obtained spin-dependent transport parameters are relatively close to values found in the literature, for Pt[15], CoFe[28,37], NiFe[12,13,38] and Ta[20]. We find a very low spin diffusion length for Ta (2 nm), and quite low spin diffusion lengths of NiFe, CoFe and Pt (5.3 nm, 6 nm and 3.8 nm respectively at room temperature). In contrast, the spin diffusion length of Co is found to be of the order of 10 nm at room temperature, smaller than in previous reports[13].

As mentioned above, the polarization and spin diffusion length are linked in most experiment. Hence, variations of the couple $(P_F, \lambda_F)$ can hence appear from one paper to another, but the spin resistance area product $\rho_F \lambda_F / (1 - P_F^2)$ should be similar. Considering low resistivities, as by extracting them from the bulk materials, can also lead to differences in the obtained spin transport parameters[22]. We emphasis also that the quite low extracted polarization for Co or NiFe might arise from the considered model, where only the bulk contribution is taken into account and not the interfacial spin filtering. This would require extra parameters and also access to the interface resistance. Finally, and while very unlikely in our studied metallic interfaces, we note that spin sink experiments might be affected by magnetic proximity effects or charge transfer in systems involving for example semi-conductors, transition metal dichalcogenides or as for graphene in contact with ferromagnetic metals or insulators, so that care has to be taken in those cases.

To conclude, we studied the absorption of pure spin currents in different materials, in order to determine their spin transport parameters. We demonstrated here that spin absorption experiments are well adapted to study ferromagnetic materials. We extracted values of spin diffusion length and polarization of several ferromagnetic element and heavy metals, at both 300 K and 10 K. This study shows that this mean of analyzing is versatile, and adapted for any material possessing a short spin diffusion length.

**ACKNOLEDGEMENT**

This work was partly supported by the LABEX Lanef ANR-10-LABX-51-01 and by the French Agence Nationale de la Recherche (ANR) through the Project ANR-SOSPIN (2013-2017).

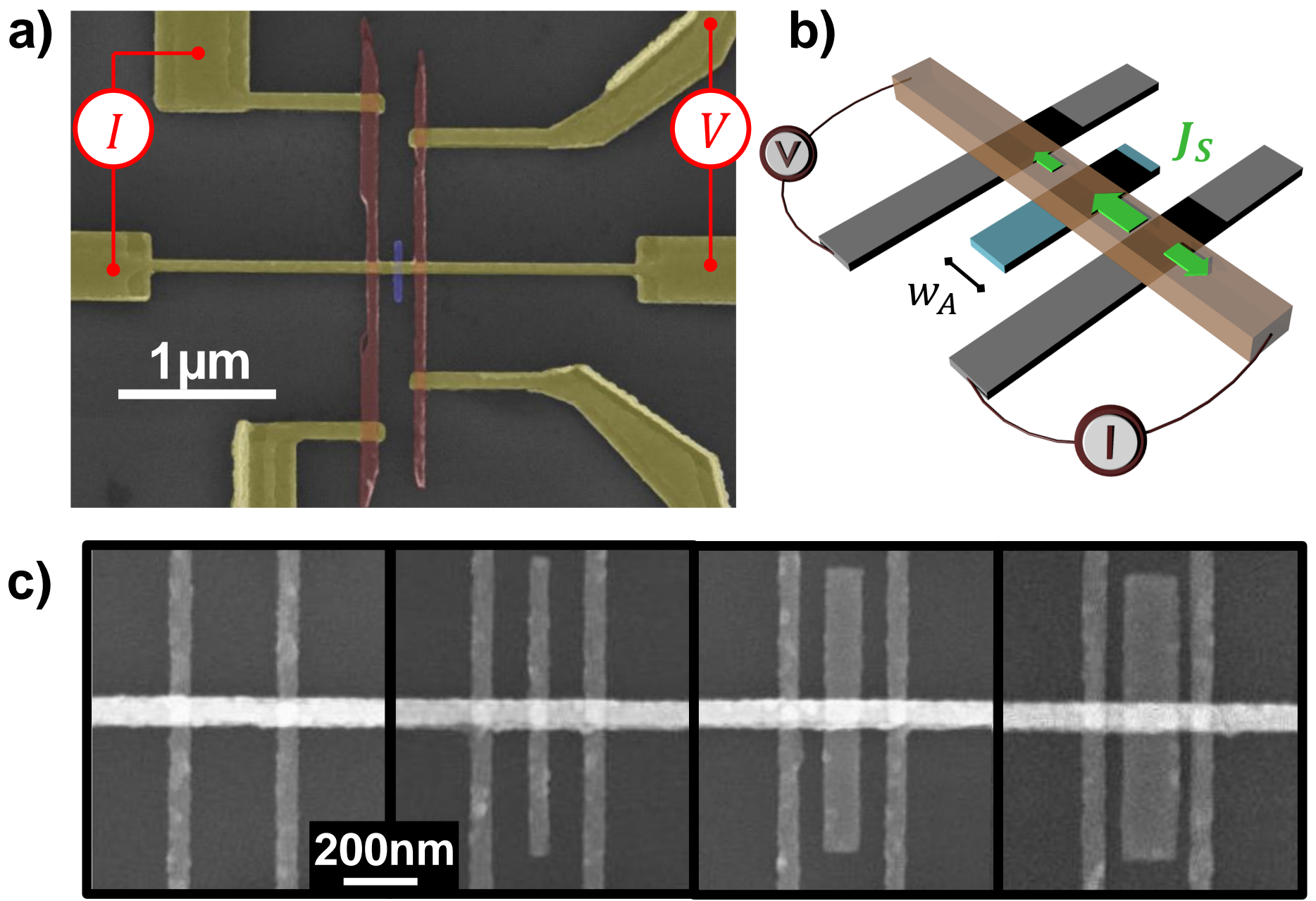

**Figure 1**
a) Colored SEM image of a spin-absorption device, displaying the measurement configuration and the constituting materials. The ferromagnetic electrodes are in red, the non-magnetic channel is yellow, and the spin-absorbing wire is blue. b) Scheme of the device, in the same measurement configuration. $w_A$ is the width of the spin-absorbing wire. The pure spin current, represented by green arrows, is partially absorbed by the central element. Due to efficient relaxation near the absorber and detector, more spin current diffuses toward this direction (here the left). c) SEM images of a set of spin-absorption devices. The left image is that of a reference lateral spin valve, whereas in the other images a spin-sink material has been inserted. These inserted wires are 50nm, 100nm and 150nm wide, respectively.

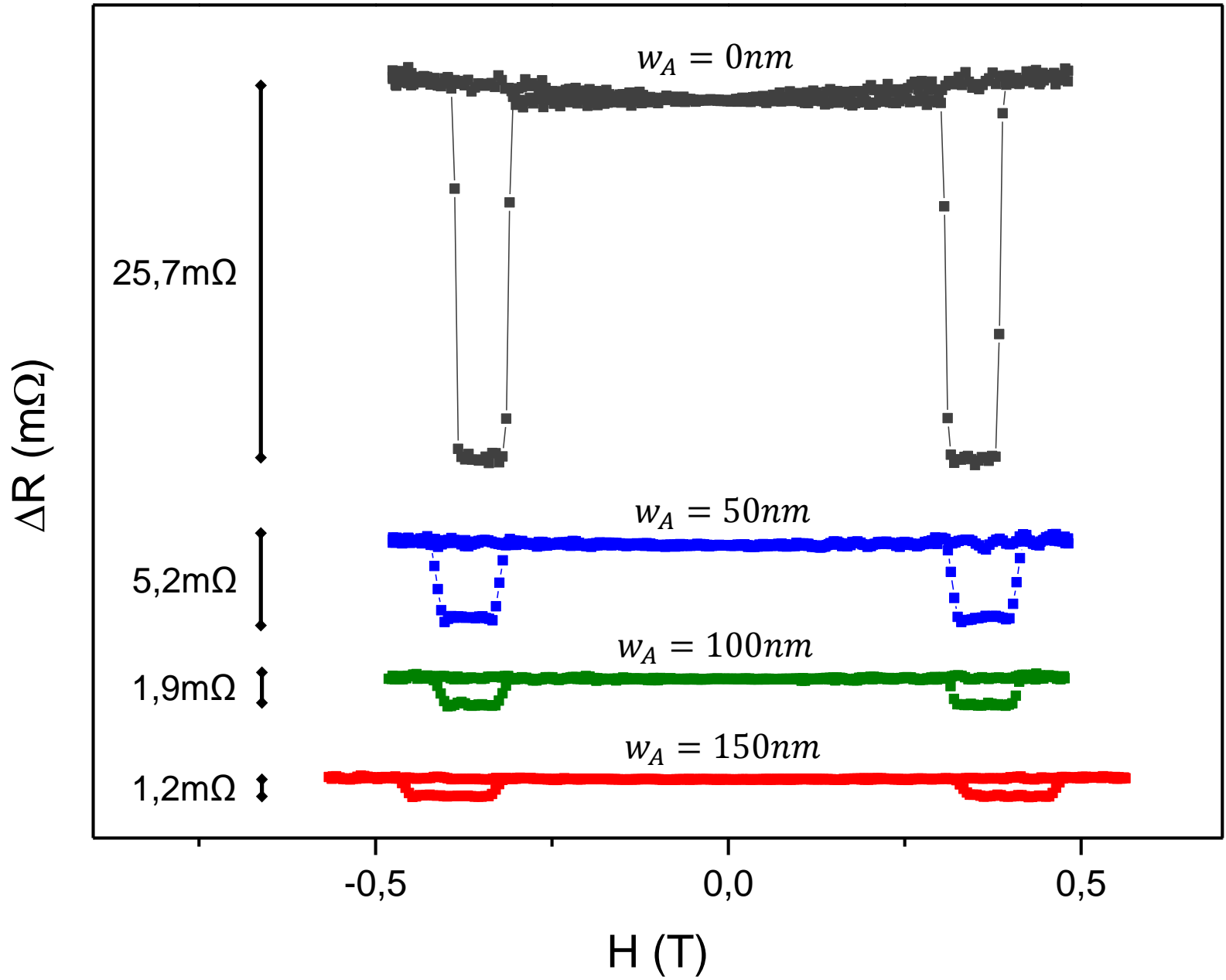

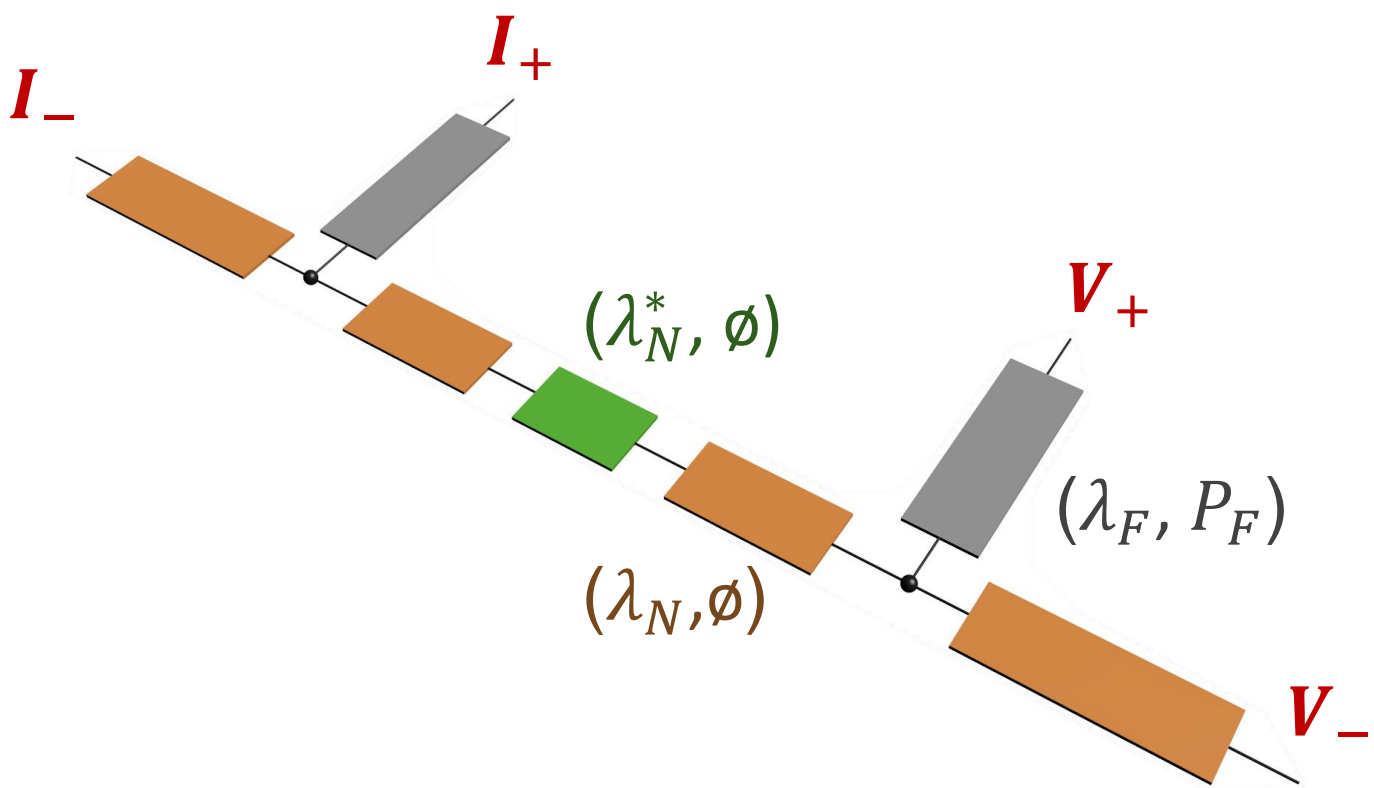

**Figure 2**
a) Non-local measurements performed at room temperature on different devices. Ferromagnetic electrodes are in CoFe alloy, and the non-magnetic channel is in Copper. Here Pt wires of different widths have been inserted between the injection and detection electrodes. One can see that the spin signal amplitude is decreasing when the width of the absorbing wire increases. Different offsets have been set for each magnetoresistance curves for graphical purpose. b) Spin resistor representation of the 1D analytical model used in this paper. The conducting channel and the ferromagnetic elements are represented respectively in brown and grey. The green spin resistor correspond to the region where the channel is in contact with the spin absorbing wire. Hence, the effective spin transport parameter $\lambda_N^*$ of this part of the channel is modified by the presence of the spin absorbing wire, as seen in equation 1.

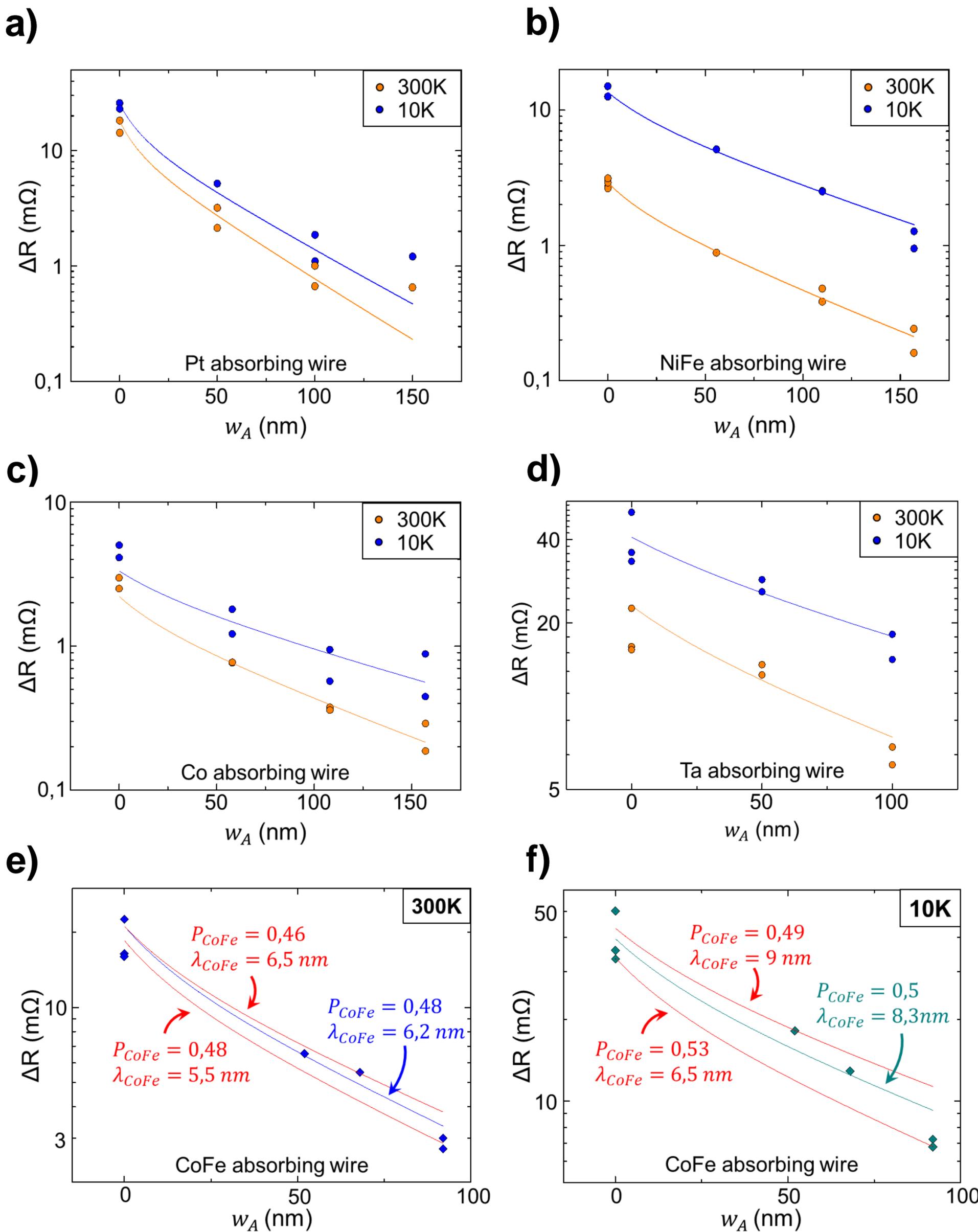

**Figure 3**
Measured and calculated spin signal amplitudes as a function of the width $w_A$ of the absorbing wire for the various absorbing materials. On figure a) (respectively b, c, d, e and f), the absorbing element is Pt (respectively NiFe, Co, Ta and CoFe). Experimental results are represented by dots on the graphs, each dot corresponding to the spin signal amplitude measured on one device. The curves correspond to the analytical expression of the one-dimensional model, enabling to extract the different material parameters. On figures a, b, c and d, both the measurements and the fitting curves are displayed at 300K (in orange) and 10K (in blue). On figure e and f, the central curve enables to see the fitting and the obtained material parameters while the red external curves show the dependence of the fitting curve for small variations of $P_F$ and $\lambda_F$. This parameter dependence enables to estimate the range of acceptable parameters value, setting the interval of confidence.